\documentclass[aps,pra,twocolumn,amssymb,eqsecnum,secnumarabic,showpacs,showkeys,floatfix]{revtex4}
\usepackage{bm}
\usepackage{amsmath}
\usepackage{natbib}

\usepackage{graphicx,amsmath,amssymb}
\usepackage{epsf,color,graphicx}

\begin{document}
\title{Generalized Hypergeometric Coherent States}
\author{T.~Appl} \author{D.~H.~Schiller} 
\affiliation{Fachbereich Physik, Universit\"{a}t Siegen, D-57068 Siegen, Germany}
\email{schiller@physik.uni-siegen.de}

\begin{abstract}
We introduce a large class of holomorphic quantum states by choosing their normalization functions to be given by generalized hypergeometric functions. We call them generalized hypergeometric states in general, and generalized hypergeometric coherent states in particular, if they allow a resolution of unity. Depending on the domain of convergence of the generalized hypergeometric functions, we distinguish generalized hypergeometric states on the plane, the open unit disk, and the unit circle. All states are eigenstates of suitably defined lowering operators. We then study their photon number statistics and phase properties as revealed by the Husimi and Pegg-Barnett phase distributions. On the basis of the generalized hypergeometric coherent states we introduce generalized hypergeometric Husimi distributions and corresponding phase distributions as well as new analytic representations of arbitrary quantum states in Bargmann and Hardy spaces.
\end{abstract}

\pacs{42.50.-p, 03.65.-w}
\keywords{coherent states; resolution of unity; Bargmann representation; phase space distributions; phase distributions.}

\maketitle

\section{Introduction} \label{S1}

Coherent states, their variants and generalizations have been extensively studied  over the last four decades. A comprehensive review of this development can be found in Refs.~\cite{klsk, fe, do}. Recently, Klauder et al. \cite{kl} have exposed a rather general method for constructing holomorphic coherent states of the form
\begin{equation}
\vert z \rangle \ = \ [\mathcal{N}(\vert z \vert^2)]^{-\frac{1}{2}}
\ \sum_{n=0}^{\infty} \ \frac{z^n}{\sqrt{\rho(n)}} \ \vert n \rangle \ ,
\label{G100}
\end{equation}
where $z$ is a complex variable, $\rho(n)$ a set of strictly positive parameters and the states $\vert n \rangle$ form an orthonormal basis (usually the Fock basis). 
The normalization function is given by the series
\begin{equation}
\mathcal{N}(\zeta) = \ \sum_{n=0}^{\infty} \frac{\zeta^n}{\rho (n)} 
\label{G105} 
\end{equation}
for $\zeta = \vert z \vert^2$; its radius of convergence determines the domain of definition of the states in Eq.~(\ref{G100}).

In this paper we consider a particular, but nevertheless very large class of holomorphic coherent states of the form (\ref{G100}), for which the normalization functions (\ref{G105}) are given by generalized hypergeometric functions 
\begin{equation}
\mathcal{N}(\zeta) = \ _{p} F_{q}(a_{1},\ldots,a_{p};b_{1},\ldots, b_{q};\zeta).
\label{G108} 
\end{equation}
We call them generalized hypergeometric states in general, and generalized hypergeometric coherent states if in addition they allow a resolution of unity with a positive weight function. They comprise many well known quantum states as particular or limiting cases as well as some of the examples constructed in Ref.~\cite{kl}.

The plan of the paper is as follows. In Section \ref{S2} we introduce the generalized hypergeometric states, which naturally fall into three classes: generalized hypergeometric states on the plane, the open unit disk, and the unit circle.
In Section \ref{S3} we discuss the generalized hypergeometric coherent states and use them to represent an arbitrary state by analytic functions either on the plane or the unit disk. 
We then show in Section \ref{S4} that all states are eigenstates of suitably defined lowering operators.
In Section \ref{S5} we consider the photon number statistics of the generalized hypergeometric states and in Section \ref{S6} their phase properties as given by the Husimi and Pegg-Barnett phase distributions. 
Generalizations of the Husimi distribution on the basis of the generalized hypergeometric coherent states and corresponding phase distributions are introduced and discussed in Section \ref{S7}.
We conclude in Section \ref{S8} with a summary of our main results.

\section{Generalized Hypergeometric States} \label{S2}

In this section we introduce the generalized hypergeometric states, classify them according to definite criteria of convergence, and indicate the constraints on the parameters on which the states depend.

We define generalized hypergeometric states (GHS) by
\begin{eqnarray}
\vert p; q; z \rangle &\equiv& \vert a_{1},\ldots, a_{p};
b_{1},\ldots, b_{q}; z \rangle _{(p;q)} \nonumber \\ &=& 
[_{p}\mathcal{N}_{q}(\vert z \vert^2)]^{-\frac{1}{2}} \
\sum_{n=0}^{\infty} \ \frac{z^n}{\sqrt{_{p}\rho_{q}(n)}} \ \vert n
\rangle , \label{G210}
\end{eqnarray}
with the strictly positive parameter functions (of the discrete variable $n$)
\begin{eqnarray}
_{p}\rho_{q}(n) &\equiv& _{p}\rho_{q}(a_{1},\ldots, a_{p};
b_{1},\ldots, b_{q};n) \nonumber \\ &=& \Gamma(n+1) \
\frac{(b_{1})_{n} \cdots (b_{q})_{n}}{(a_{1})_{n} \cdots (a_{p})_{n}} , 
\label{G230}
\end{eqnarray}
where $(a)_{n} = \Gamma(a+n)/\Gamma(a)$ is the Pochhammer symbol. 
The normalization function is then given by the generalized hypergeometric function
\begin{eqnarray}
_{p} \mathcal{N}_{q}(\zeta) &=& \ _{p} F_{q}(a_{1},\ldots,
a_{p};b_{1},\ldots, b_{q};\zeta) \nonumber \\ &=& \sum_{n=0}^{\infty}
\frac{(a_{1})_{n} \cdots (a_{p})_{n}}{(b_{1})_{n} \cdots (b_{q})_{n}} \ \frac{\zeta^n}{n!} \label{G200}
\end{eqnarray}
evaluated at $\zeta = \vert z \vert^2$. The GHS depend on the complex variable $z$ and on the two sets of numerator $(a_{1},\ldots,a_{p})$ and denominator $(b_{1},\ldots, b_{q})$ parameters.  
The non-negative integers $p$ and $q$ assume the value $p=0$ ($q=0$) if there are no numerator (denominator) parameters; this will be indicated by a dot, e.g. $\vert 1;0;z \rangle = \vert a; \cdot \thinspace ; z \rangle _{(1;0)}$. 
Since the parameter functions $_{p}\rho_{q}(n)$ are the defining quantities, some of their particular properties become ``global'' and hold true for the GHS and all quantities derived on their basis. This will by exemplified repeatedly in the following. As a first property we mention the symmetric dependence on the numerator and, separately, the denominator parameters. A second property is related with the factorized structure of $_{p}\rho_{q}(n)$: if $l$ numerator and $l$ denominator parameters coalesce ($a_{i} = b_{j}$), they cancel and the state $\vert p; q; z \rangle$ reduces to the lower-order state $\vert p-l; q-l; z \rangle$ of the remaining parameters. Thus, the conventional coherent states (CS) result for $p=q=0$ or, more generally, for $p=q \neq 0$ and all parameters coalescing.

The normalization function also determines the scalar
product of two GHS with identical parameter sets
\begin{equation}
\langle p;q;z \vert p;q;z' \rangle \ = \frac{_{p}
\mathcal{N}_{q}(z^{*} z')}{\sqrt{_{p} \mathcal{N}_{q}(\vert z \vert^2) 
\ _{p} \mathcal{N}_{q}(\vert z' \vert^2)}} . \label{G237}
\end{equation}
It follows that the GHS are normalized, but not orthogonal. The scalar product is well defined if the generalized hypergeometric functions involved converge. The function $_{p}F_{q}(a_{1},\ldots,a_{p};b_{1},\ldots,b_{q};\zeta)$ converges in the following cases for:
\begin{subequations}
\label{allG233s}
\begin{eqnarray}
{\text{any finite}} \ \zeta \ &{\text{if}}& \ p < q+1; \label{G233a} \\
\vert \zeta \vert < 1 \ &{\text{if}}& \ p=q+1; \label{G233b} \\ \vert \zeta
\vert = 1 \ &{\text{if}}& \ p=q+1, \ \eta<0; \label{G233c} \\ \vert
\zeta \vert = 1, \ \zeta \neq 1 \ &{\text{if}}& \ p=q+1, \ 0 \le \eta < 1, \label{G233d}
\end{eqnarray}
\end{subequations}
where $\eta \equiv \text{Re}(\sum_{j=1}^{p}a_{j} - \sum_{j=1}^{q}b_{j})$; in all other cases it diverges \cite{ma}. According to the three cases of unconditional convergence, Eqs.~(\ref{G233a}) to (\ref{G233c}), we obtain generalized hypergeometric states on the plane, the open unit disk, and the unit circle, respectively. These states depend analytically on the complex variable $z$ (disregarding the normalization function) and thus belong to the very large class of holomorphic quantum states.
Besides these normalized states we shall consider also unnormalizable states on the unit circle violating the condition $\eta < 0$.

We conclude this section by reviewing the constraints on the parameters $a_{1} , \ldots, a_{p}$ and $b_{1} , \ldots, b_{q}$.
First of all, we exclude zero or negative integer values for $a_{1}, \ldots, a_{p}$ (leading to undefined parameter functions $_{p}\rho_{q}$) and for $b_{1}, \ldots, b_{q}$ (leading to undefined normalization functions $_{p}F_{q}$). Next we require the parameter functions $_{p}\rho_{q}(n)$ to be real and strictly positive, leading to the
conditions
\begin{equation}
\frac{(b_{1}+n) \cdots (b_{q}+n)}{(a_{1}+n) \cdots (a_{p}+n)} > 0 \ \ {\text{for all}} \ \ n = 0, 1, 2,\ldots \ ,     \label{G221}
\end{equation}
which follow from the obvious recurrence relation
\begin{equation}
\frac{_{p}\rho_{q}(n+1)}{(n+1)!} \ = \ \frac{_{p}\rho_{q}(n)}{n!} \
\frac{(b_{1}+n) \cdots (b_{q}+n)}{(a_{1}+n) \cdots (a_{p}+n)} \label{G222}
\end{equation}
with seed value $_{p}\rho_{q}(0) = 1$. The conditions
(\ref{G221}) are satisfied for all $n$, if: (i) all parameters are
real and positive, (ii) an even number of parameters is real and
negative with pairwise the same negative integer parts, 
(iii) an even number of numerator parameters and/or of denominator
parameters are pairwise complex conjugate to each other.
If cases (ii) and/or (iii) occur, the remaining parameters are supposed to be real and positive.

\section{Generalized Hypergeometric Coherent States} \label{S3}

In this section we introduce generalized hypergeometric coherent states (GHCS) as a set of generalized hypergeometric states (GHS) which is (over)complete and allows a resolution of unity with a non-negative weight function. The resolution of unity is then used on the one hand to determine the corresponding weight functions and on the other hand to introduce new analytic representations for arbitrary quantum states.

We consider first the GHCS on the plane and on the unit disk related,
respectively, with the convergence criteria (\ref{G233a}) and (\ref{G233b}). In these cases the resolution of unity ($\hat 1$) takes the form
\begin{equation}
\frac{1}{\pi} \ \int d^2 z \ _{p}w_{q}(\vert z \vert^2) \ \vert p;
q; z \rangle \ \langle p; q; z \vert \ = \ \hat 1 , \\ \label{G280}
\end{equation}
with a positive weight function $_{p}w_{q}(x)$
depending on $x=\vert z \vert^2$; the integration is over the complex plane or unit disk, as appropriate. 
Introducing the states (\ref{G210}) into Eq.~(\ref{G280}) and performing the angular integration, the following conditions result:
\begin{equation}
\int\limits_{0}^{R} dx \ x^n \ _{p} \tilde w_{q}(x) \ = \ _{p}\rho_{q}(n) , \quad n=0,1,2, \ldots , \label{G281}
\end{equation}
where $_{p} \tilde w_{q}(x) \equiv \ _{p}w_{q}(x) / _{p}\mathcal{N}_{q}(x)$ and the upper limit of the integral is $R=\infty$ ($R=1$) for the GHCS on the plane (disk). Correspondingly, the unknown distribution $_{p} \tilde w_{q}(x)$ is given by the solution of a Stieltjes (Hausdorff) moment problem with the moments given in Eq.~(\ref{G230}). As shown by Klauder et al. \cite{kl}, the solution can be obtained by using Mellin transform techniques. Thus, replacing in Eq.~(\ref{G281}) the discrete variable $n$ by the complex variable $(s-1)$, 
the distribution $_{p} \tilde w_{q}(x)$ and the parameter function $_{p}\rho_{q}(s-1)$ become a Mellin transform related pair. There are 
well known references where such pairs are tabulated \cite{ma, pr, ob}. Thus, using Ref.~\cite[p. 303, formula (37)]{ma} or Ref.~\cite[p. 728, formula (8.4.51.9)]{pr}, we find in terms of the Meijer $G$-function
\begin{eqnarray}
\int\limits_{0}^{R} dx \ x^{s-1} \ G_{p , \ q+1}^{q+1 , \ 0} \bigg(
x \bigg\vert \begin{array}{l}a_{1}-1,\ldots, a_{p}-1 \quad \\[2pt] b_{1}-1,\ldots,
b_{q}-1, 0\end{array} \bigg) \nonumber \\ = \ \Gamma(s) \
\frac{\Gamma(s+b_{1}-1) \cdots \Gamma(s+b_{q}-1)}{\Gamma(s+a_{1}-1) \cdots \Gamma(s+a_{p}-1)} \ , \label{G290}
\end{eqnarray}
where the r.h.s.\ is the $s$-dependend part of $_{p}\rho_{q}(s-1)$. 
It follows that the weight function in Eq.~(\ref{G280}) is given by
\begin{eqnarray}
_{p}w_{q}(x) &\equiv& _{p}w_{q}(a_{1},\ldots, a_{p};b_{1},\ldots,
b_{q};x) \nonumber \\ &=& \frac{\Gamma(a_{1}) \cdots \Gamma(a_{p})}{\Gamma(b_{1}) \cdots \Gamma(b_{q})} \ _{p}F_{q}(a_{1},\ldots,
a_{p};b_{1},\ldots, b_{q};x) \nonumber \\ & & \times \ G_{p,q+1}^{q+1,0} \bigg( x \bigg\vert \begin{array}{l}a_{1}-1,\ldots, a_{p}-1 \quad \\[2pt] b_{1}-1,\ldots,
b_{q}-1, 0\end{array} \bigg) . \label{G294}
\end{eqnarray}
The Mellin transform (\ref{G290}) is valid if either
\begin{subequations}
\label{allG297s}
\begin{eqnarray}
&& p < q+1 , \quad {\text {or}} \label{G297a} \\
&& p = q + 1 , \ \eta > 1 \label{G297b}
\end{eqnarray}
\end{subequations}
are satisfied. It follows that $_{p}w_{q}(x)$ exists for all GHS on the plane, but only for those with $\eta > 1$ on the disk. The solution found, however, need not be positive. 
The positivity of $_{p}w_{q}(x)$ must be checked in each particular case and may result in additional constraints on the parameters. As was the case for the states, if $l$ numerator parameters coalesce with $l$ denominator parameters, the weight function $_{p}w_{q}(x)$ reduces to the lower-order weight function $_{p-l}w_{q-l}(x)$ of the remaining parameters, due to a well known property of the $_{p}F_{q}$ and $G$ functions. On the other hand, this can be seen also as a consequence of the defining condition (\ref{G281}).

The resolution of unity, Eq.~(\ref{G280}), can be used to introduce a new representation by sandwiching it between two arbitrary state vectors, $\langle \Phi \vert$ and $\vert \Psi \rangle$, and writing the resulting scalar product in the form
\begin{equation}
\langle \Phi \vert \Psi \rangle \ = \ \frac{1}{\pi} \int d^2 z \ _{p}\Phi_{q}^{*}(z) \ _{p}\Psi_{q}(z) , \label{G298}
\end{equation}
with the wave functions in the GHCS basis defined according to
\begin{eqnarray}
_{p} \Psi_{q}(z) &=& \sqrt{ _{p} w_{q}(\vert z \vert^2)} \ \langle p; q; z \vert \Psi \rangle \nonumber \\
&=& \Big[ \frac{_{p}w_{q}(\vert z \vert^2)}{ _{p}\mathcal{N}_{q}(\vert z \vert^2)} \Big]^{\frac{1}{2}} \ \sum_{n=0}^{\infty} \frac{(z^*)^n}{\sqrt{_{p}\rho_{q}(n)}} \ \langle n \vert \Psi \rangle . \label{G299}
\end{eqnarray}
We call this the GHCS representation of the state $\vert \Psi \rangle$.
The series appearing in Eq.~(\ref{G299}) defines an entire analytic function of $\zeta = z^{*}$, which may be regarded as another representation. We call it the generalized hypergeometric analytic representation of $\vert \Psi \rangle$:
\begin{equation}
_{p} \tilde \Psi_{q}(\zeta ) \ = \ \sum_{n=0}^{\infty} 
\frac{\zeta ^n}{\sqrt{_{p}\rho_{q}(n)}} \ \langle n \vert \Psi \rangle . \label{G302}
\end{equation}
The scalar product (\ref{G298}) then writes as
\begin{equation}
\langle \Phi \vert \Psi \rangle \ = \ \int \ _{p} d \mu_{q}(\zeta ) \ \ _{p}{\tilde \Phi}_{q}^{*}(\zeta ) \ _{p}{\tilde \Psi}_{q}(\zeta ) , \label{G303}
\end{equation}
with the measure given by
\begin{equation}
_{p} d \mu_{q}(\zeta ) \ = \ \frac{d^2 \zeta}{\pi} \ \frac{_{p}w_{q}(\vert \zeta  \vert^2)}{_{p}{\mathcal N}_{q}(\vert \zeta \vert^2)} \ . \label{G305}
\end{equation}
Note that the analytic representations and the measures depend explicitly also on the parameter sets ($a_{1},\ldots, a_{p}$) and ($b_{1},\ldots,b_{q}$). For a given state $\vert \Psi \rangle$ we then obtain infinitely many analytic representations on the plane and unit disk beeing, respectively, elements of Bargmann and Hardy spaces with measure (\ref{G305}). The usual Bargmann representation based on the coherent states is recovered for $p=q=0$. The simplest Hardy space representation is based on the coherent phase states given in Eq.~(\ref{G260}) below \cite{vou2}.

Next we consider the GHS on the unit circle and write $z={\text{e}}^{i \varphi}$. We distinguish two types of states, denoted by $\vert p; q; {\text{e}}^{i \varphi} \rangle$ and $\vert p; q; \varphi \rangle$ depending on whether the convergence criterion (\ref{G233c}) is satisfied or not. The normalized states ($\eta<0$) are defined by
\begin{equation}
\vert p; q; {\text{e}}^{i \varphi} \rangle \ = \ 
\frac{1}{\sqrt{_{p}{\mathcal N}_{q}(1)}} \sum_{n=0}^{\infty} 
\frac{{\text{e}}^{i n \varphi}}{\sqrt{ _{p}\rho_{q}(n)}} \ \vert n \rangle \label{G307}
\end{equation}
and the unnormalizable states ($\eta \ge 0$) by
\begin{equation}
\vert p; q; \varphi \rangle \ = \ \frac{1}{\sqrt{2 \pi}} \sum_{n=0}^{\infty} \frac{{\text{e}}^{i n \varphi}}{\sqrt{ _{p}\rho_{q}(n)}} \ \vert n \rangle . \label{G308}
\end{equation}
The normalization function in Eq.~(\ref{G307}) is a constant given by the generalized hypergeometric function of unit argument. 
These states do not yield a resolution of unity, as the resulting Hausdorff moment problem
\begin{equation}
\int\limits_{0}^{2 \pi} d \varphi \ _{p} \tilde w_{q}(\varphi) \ {\text{e}}^{i(n-m)\varphi} \ = \ _{p}\rho_{q}(n) \ \delta_{n,m} \label{G309}
\end{equation}
for the sought for function $_{p} \tilde w_{q}(\varphi)$ has no solution: Eq.~(\ref{G309}) implies the vanishing of all Fourier components of the function $_{p} \tilde w_{q}(\varphi)$ for $n \neq m$, thus reducing it to a constant which cannot match the values on the r.h.s. for all $n = m$. The lowest possible states 
$\vert 1; 0; {\text{e}}^{i \varphi}\rangle = 
\vert a; \cdot \thinspace ; {\text{e}}^{i \varphi}\rangle _{(1;0)}$ 
are not allowed, as the only parameter $a$ cannot satisfy the conditions (\ref{G233c}) and (\ref{G221}) simultaneously. The lowest normalizable states are $\vert 2; 1; {\text{e}}^{i \varphi}\rangle = 
\vert a_{1}, a_{2}; b; {\text{e}}^{i \varphi}\rangle _{(2;1)}$ for 
$\eta = a_{1} + a_{2} - b < 0$ in addition to the constraints (\ref{G221}); 
their normalization constant is given by the Gaussian hypergeometric function of unit argument, $_{2}F_{1}(a_{1},a_{2};b;1)= \Gamma(b)\Gamma(b-a_{1}-a_{2})/[\Gamma(b-a_{1})\Gamma(b-a_{2})]$. 
On the other hand, the simplest unnormalizable states are $\vert 1; 0; \varphi \rangle = \vert a; \cdot \thinspace; \varphi \rangle _{(1;0)}$, which in the limit $a \rightarrow 1$ yield the well known phase states \cite{lo}
\begin{equation}
\vert a; \cdot \thinspace; \varphi \rangle _{(1;0)} \mid_{a \rightarrow 1} \ = \ 
\vert \varphi \rangle \ \equiv \ \frac{1}{\sqrt{2\pi}} \ \sum_{n=0}^{\infty} \ {\text{e}}^{i n \varphi} \ \vert n \rangle . \label{G261}
\end{equation}
These states, corresponding to 
$_{1}\rho_{0}(a; \cdot \thinspace ; n) \mid_{a \rightarrow 1} = 1$ for all $n$, are the only ones yielding a resolution of unity with $_{1} \tilde w_{0}= 1/(2 \pi)$  according to Eq.~(\ref{G309}).

We conclude this section by giving examples of GHCS for particular values of $p$ and $q$. We quote for each state the parameter, the normalization and the weight functions. The complex variable $z$ is denoted by $\alpha$, $\vert \alpha \vert < \infty$, for the GHCS on the plane ($p<q+1$) and by $\epsilon$, $\vert \epsilon \vert < 1$, for the GHCS on the unit disk ($p=q+1$).

(a) The conventional coherent states 
$\vert \alpha \rangle = \vert 0; 0; \alpha \rangle$ correspond to $p=q=0$, in
which case $_{0}\rho_{0}(n)=n!$, $_{0}F_{0}(\zeta) \ = \ {\text{e}}^{\zeta}$ and $_{0}w_{0}(\vert \alpha \vert^2)=1$.

(b) The states 
$\vert 0; 1; \alpha \rangle = \vert \cdot \thinspace ; b; \alpha \rangle _{(0;1)}$,  
related with $_{0}\rho_{1}(\cdot \thinspace ; b; n) = n! (b)_{n}$ for $b>0$ due to Eq.~(\ref{G221}), have the normalization function
\begin{equation}
_{0} F_{1}(\cdot \thinspace ; b; \zeta) \ = \ \Gamma(b) \ \zeta^{\frac{1-b}{2}} \ I_{b-1}(2 \sqrt{\zeta})  \label{G410}
\end{equation}
and the weight function (\cite[p. 196, formula (5.39)]{ob})
\begin{equation}
_{0}w_{1}(\cdot \thinspace ; b; \vert \alpha \vert^2) = 2 \ I_{b-1}(2 \vert \alpha
\vert) \ K_{b-1}(2 \vert \alpha \vert) , \label{G430}
\end{equation}
where $I_{\nu}(x)$ and $K_{\nu}(x)$ are the modified Bessel
functions. Since they are real and positive when $\nu>-1$ and $x>0$
\cite{as}, the weight function $_{0}w_{1}$ is positive for $b>0$.

(c) For the states 
$\vert 1; 1; \alpha \rangle = \vert a; b; \alpha \rangle _{(1;1)}$ 
with  $_{1}\rho_{1}(a; b; n)=n!(b)_{n}/(a)_{n}$, the positivity
constraints (\ref{G221}) require $a$ and $b$ to be either both
positive or both negative with equal negative integer parts. The
normalization function is given by Kummer's confluent hypergeometric
function $_{1}F_{1}(a;b;\zeta)$ and the weight function by (\cite[p. 716, formula (9)]{pr})
\begin{eqnarray}
_{1}w_{1}(a;b;\vert \alpha \vert^2) &=& \frac{\Gamma(a)}{\Gamma(b)}
\ _{1}F_{1}(a; b; \vert \alpha \vert^2) \ \times \nonumber \\ & &
{\text{e}}^{- \vert \alpha \vert^2} \ \Psi(a-b;2-b;\vert \alpha
\vert^2) , \  \    \label{G460}
\end{eqnarray}
with $\Psi(a;b;\zeta)$ the Tricomi confluent hypergeometric function. For $a=b$ the above results reduce to those for the coherent states $\vert 0;0;\alpha \rangle$ in example (a).

(d) The states 
$\vert 1; 0; \epsilon \rangle = \vert a; \cdot \thinspace; \epsilon \rangle _{(1;0)}$ 
originate from $_{1}\rho_{0}(a; \cdot \thinspace ; n) = n! / (a)_{n}$, with $a>0$ due to Eq.~(\ref{G221}). The normalization function is given by
\begin{equation}
_{1} F_{0}(a; \cdot \thinspace ; \zeta ) \ = \ (1 - \zeta )^{-a}   \label{G300}
\end{equation}
and the weight function by (\cite[p. 195, formula (5.35)]{ob})
\begin{equation}
_{1}w_{0}(a; \cdot \thinspace ; \vert \epsilon \vert^2) = \frac{a-1}{(1- \vert \epsilon \vert^2)^2} \  , \label{G330}
\end{equation}
with $a>1$ due to Eq.~(\ref{G297b}), in which case also $_{1}w_{0}>0$.

The states $\vert a; \cdot \thinspace; \epsilon \rangle _{(1;1)}$ for $a=2k$ and Bargmann index $k=\frac{1}{2},1,\frac{3}{2},\ldots$ yield the Perelomov coherent states of the discrete series of the SU(1,1) representations, and for arbitrary non-negative $k$ those of its universal covering group \cite{pe1,pe2}. Of particular interest are the states for $k=\frac{1}{2}$, also known as coherent phase states \cite{sh}
\begin{equation}
\vert a; \cdot \thinspace; \epsilon \rangle _{(1;0)} \mid_{a \rightarrow 1} \ = \ 
\vert \epsilon \rangle \ \equiv  \ \sqrt{1 - \vert \epsilon \vert^2} \
\sum_{n=0}^{\infty} \ \epsilon^n \ \vert n \rangle \ . \label{G260}
\end{equation}
Their weight function (\ref{G330}) vanishes and the resolution of unity is applicable only as a limiting equation \cite{vou1}.

(e) For the states 
$\vert 2; 1; \epsilon \rangle = \vert a_{1}, a_{2}; b; \epsilon \rangle _{(2;1)}$ 
with $_{2}\rho_{1}(a_{1}, a_{2}; b;n)=n!(b)_{n}/[(a_{1})_{n}(a_{2})_{n}]$,
the constraints (\ref{G221}) allow not only positive parameter values, but also pairs of negative-valued parameters with equal negative integer parts as well as a complex-valued pair ($a_{2} = a_{1}^{*}$). 
The normalization function is given by the Gaussian hypergeometric function $_{2}F_{1}(a_{1},a_{2};b; \zeta)$ and 
the weight function by (\cite[p. 198, formula (5.50)]{ob})
\begin{eqnarray}
_{2}w_{1}(a_{1},a_{2};b;\vert \epsilon \vert^2) =
\frac{\Gamma(a_{1}) \ \Gamma(a_{2})}{\Gamma(b) \
\Gamma(a_{1}+a_{2}-b-1)} \times \nonumber \\ (1 - \vert \epsilon
\vert^2)^{a_{1}+a_{2}-b-2} \ _{2}F_{1}(a_{1},a_{2};b;\vert \epsilon
\vert^2) \times \nonumber \\
_{2}F_{1}(a_{2}-b,a_{1}-b;a_{1}+a_{2}-b-1;1 - \vert \epsilon
\vert^2) , \  \  \label{G385}
\end{eqnarray}
with $a_{1}+a_{2}-b>1$ due to Eq.~(\ref{G297b}). It is not clear if this condition together with the constraints (\ref{G221}) are sufficient to guarantee the positivity of $_{2}w_{1}$ in all cases. For coalescing parameters ($a_{1}=b$ or $a_{2}=b$) we recover the results of the previous example (d) for the states 
$\vert 1; 0; \epsilon \rangle$.

\section{Generalized Hypergeometric States as Eigenstates of Lowering Operators} \label{S4}

In this section we introduce ladder operators of raising and 
lowering type, the latter having as eigenstates the generalized
hypergeometric states introduced in Section \ref{S2}. These operators
generalize the creation and annihilation operators of the harmonic oscillator in the case of the GH(C)S on the plane, and the Susskind-Glogower exponential phase
operators in the case of the GH(C)S on the unit disk and circle.

We define generalized hypergeometric lowering and raising operators
by
\begin{subequations}
\label{allG610s}
\begin{eqnarray}
_{p}\hat U_{q} \ = \ \sum_{n=0}^{\infty} \ _{p}f_{q}(n) \ \vert n \rangle \langle n+1 \vert  , \label{G610a} \\ _{p}\hat
U_{q}^{\dagger} \ = \ \sum_{n=0}^{\infty} \ _{p}f_{q}(n) \ \vert n+1 \rangle \langle n \vert , \label{G610b}
\end{eqnarray}
\end{subequations}
where the coefficients $_{p}f_{q}(n)$  are given by
\begin{eqnarray}
_{p}f_{q}(n) &\equiv& _{p}f_{q}(a_{1}, \ldots, a_{p}; b_{1}, \ldots, b_{q}; n) \nonumber \\ &=& \sqrt{(n+1) \frac{(n+b_{1}) \cdots (n+b_{q})}{(n+a_{1}) \cdots (n+a_{p})}} , \label{G160}
\end{eqnarray}
with $_{0}f_{0}(n)=\sqrt{n+1}$ and all products in the numerator (denominator) replaced by $1$ if $q=0 \ (p=0)$. It is very satisfactory that these coefficients are real and positive due to the same constraints (\ref{G221}) already observed. The ladder operators depend explicitly on the two parameter sets, e.g.  
$ _{p}\hat U_{q} \ = \ _{p}\hat U_{q}(a_{1}, \ldots, a_{p}; b_{1}, \ldots, b_{q})$. 
Due to the product structure of $_{p}f_{q}(n)$, if $l$ numerator parameters coalesce with $l$ denominator parameters, the operator $ _{p}\hat U_{q}$ reduces to the lower-order operator $_{p-l}\hat U_{q-l}$ of the remaining parameters.
The ladder operators obey the noncanonical commutation relation 
\begin{equation}
[  _{p}\hat U_{q} ,\ _{p}\hat U_{q}^{\dagger}  ] \ = \ \sum_{n=0}^{\infty} \ \Big( \ _{p}f_{q}(n)^2 \ - \ _{p}f_{q}(n-1)^2 \Big) \ \vert n \rangle \langle n \vert , \label{G611}
\end{equation}
where $ _{p}f_{q}(-1) = 0$ by definition, but also by Eq.~(\ref{G160}).

The action of the ladder operators on the Fock basis is given by
\begin{subequations}
\label{allG620s}
\begin{eqnarray}
_{p}\hat U_{q} \ \vert n \rangle  \ &=& \  _{p}f_{q}(n-1) \ \vert n-1 \rangle , 
\label{G620a} \\ 
_{p}\hat U_{q}^{\dagger} \ \vert n \rangle \ &=& \  _{p}f_{q}(n) \ \vert n+1 \rangle , 
\label{G620b}
\end{eqnarray}
\end{subequations}
justifying their name as lowering and raising operators, respectively.
It is then easy to show that the GH(C)S are eigenstates of the lowering operator 
\begin{equation}
_{p}\hat U_{q} \ \vert p; q; z \rangle \ = \ z \ \vert p; q; z \rangle  , 
\label{G640}
\end{equation}
with complex eigenvalue $z$. Here the parameter functions $_{p}\rho_{q}(n)$ defining the eigenstates and the coefficients $_{p}f_{q}(n)$ defining the lowering operator are related by
\begin{equation}
_{p}\rho_{q}(n) \ = \ \Big( \ _{p}f_{q}(0) \ _{p}f_{q}(1) \ _{p}f_{q}(2) \cdots \ _{p}f_{q}(n-1) \Big)^2 .
\label{G645}
\end{equation}
Seen from this perspective the generalized hypergeometric (coherent) states are also instancies of the very large class of so-called nonlinear coherent (or f-coherent) states \cite{mfv, mmsz, do}. These are eigenstates of lowering operators of the type (\ref{G610a}) with $_{p}f_{q}(n)$ replaced by $\sqrt{n+1} \ f(n)$, where the f-parameters $f(n)$ are quite arbitrary. In our case $f(n)$ is given by Eq.~(\ref{G160}) without the factor $\sqrt{n+1}$.

For $p<q+1$ we denote 
$\ _{p} \hat A_{q} \ = \ _{p} \hat U_{q}$ and
$\ _{p} \hat A_{q}^{\dagger} \ = \ _{p} \hat U_{q}^{\dagger}$, which
may be considered as a generalization of the usual annihilation ($\hat a$) and creation ($\hat a^{\dagger}$) operators of the harmonic oscillator. In
particular, we have
\begin{equation}
_{0} \hat A_{0} \ = \ \hat a \ \equiv \ \sum_{n=0}^{\infty} \ \sqrt{n+1}
\ \vert n \rangle \langle n+1 \vert \  . \label{G660}
\end{equation}
It is then natural to call $_{p} \hat A_{q}$ ($_{p} \hat
A_{q}^{\dagger}$) the generalized hypergeometric annihilation
(creation) operators. The eigenstates of $_{p} \hat A_{q}$ are the
GH(C)S on the plane, $\vert p; q; \alpha \rangle$, with complex
eigenvalue $\alpha$, $\vert \alpha \vert < \infty$. They may be considered as generalizations of the conventional coherent states $\vert \alpha \rangle$, which are eigenstates of $\hat a$.

For $p=q+1$ we write  
$\ _{p} \hat E_{q} \ = \ _{p} \hat U_{q}$ and
$\ _{p} \hat E_{q}^{\dagger} \ = \ _{p} \hat U_{q}^{\dagger}$.
The eigenstates of $\ _{p} \hat E_{q}$ are: 
(a) the GH(C)S on the unit disk, $\vert p; q; \epsilon \rangle$, with complex eigenvalue $\epsilon$, $\vert \epsilon \vert < 1$; 
(b) the GHS on the unit circle with eigenvalue ${\text{e}}^{i \varphi}$ for both, the normalized states $\vert p; q; {\text{e}}^{i \varphi} \rangle$, Eq.~(\ref{G307}), and the unnormalizable states $\vert p; q; \varphi \rangle$, Eq.~(\ref{G308}). 
This is similar to the Susskind-Glogower \cite{sg} exponential phase operator 
$\hat E$, having as eigenstates the coherent phase states (\ref{G260}) on the unit disk and the phase states (\ref{G261}) on the unit circle. Now the operator $\hat E$ is a limiting case of the operator $_{1} \hat E_{0}(a)$ for $a \rightarrow 1$:
\begin{equation}
_{1} \hat E_{0}(a) \mid_{a \rightarrow 1} \  = \hat E \  \equiv \  
\sum_{n=0}^{\infty} \vert n \rangle \langle n+1 \vert \  .
\label{G661}
\end{equation} 
This correlates with the limit for the corresponding eigenstates in Eqs.~(\ref{G260}) and (\ref{G261}). It is then justified to consider the operators 
$\ _{p} \hat E_{q}$ and their eigenstates, the GHS on the unit disk and unit circle, as generalizations of the Susskind-Glogower exponential phase operator $\hat E$ and its eigenstates, the coherent phase states and the phase states, respectively.
For this reason we call $_{p} \hat E_{q}$ and $_{p} \hat E_{q}^{\dagger}$ the generalized hypergeometric exponential phase operators.

Given the nonhermitian ladder operators we may define hermitian combinations by
$ _{p} \hat Q_{q} = (_{p} \hat A_{q}^{\dagger} \ + \ _{p} \hat A_{q})/\sqrt{2}$ and  
$ _{p} \hat P_{q} = i (_{p} \hat A_{q}^{\dagger} \ - \ _{p} \hat A_{q})/\sqrt{2}$, 
and similarly by 
$ _{p} \hat C_{q} = (_{p} \hat E_{q}^{\dagger} \ + \ _{p} \hat E_{q})/2$ and 
$ _{p} \hat S_{q} = i (_{p} \hat E_{q}^{\dagger} \ - \ _{p} \hat E_{q})/2$.
Our notation suggests to ``interpret'' $_{p}\hat Q_{q}$ ($_{p}\hat P_{q}$) as generalized hypergeometric coordinate (momentum) operators, and $_{p}\hat C_{q}$ ($_{p}\hat S_{q}$) as generalized
hypergeometric cosine (sine) operators (\'a la Susskind-Glogower).
However, these operators do not satisfy the usual commutation relations and we have not verified that they have the right spectrum. Their study is beyond the scope of this paper.

\section{Photon Number Statistics} \label{S5}

The generalized hypergeometric (coherent) states lead to classical as well as nonclassical behaviour in the photon number statistics. The considerations in this
section do not depend on the existence of a weight function and apply, therefore, to both GHS and GHCS.

The GH(C)S defined in Eq.~(\ref{G210}) have the Fock representation
\begin{equation}
\langle n \vert p; q; z \rangle \ = \ 
\frac{z^n}{\sqrt{_{p}\rho_{q}(n) \ _{p}\mathcal{N}_{q}(\vert z \vert^2)}} \ , \label{G125}
\end{equation}
from which the photon number distribution follows
\begin{equation}
P_{\vert p; q; z \rangle}(n) \ = \ \frac{1}{_{p}\rho_{q}(n)} \
\frac{x^n}{_{p}\mathcal{N}_{q}(x)} \ \ \ \ \ (x=\vert z \vert^2) .  \label{G130}
\end{equation}
The factorial moments 
$n^{(k)} = \langle n (n-1) \cdots (n-k+1) \rangle$
are then easily calculated in terms of the derivatives of the normalization function
\begin{equation}
_{p}n_{q}^{(k)}(x) \ = \
\frac{x^k}{_{p}\mathcal{N}_{q}(x)} \ \frac{d^k}{dx^k} \
_{p}\mathcal{N}_{q}(x) \ .
\end{equation}
In our case 
$_{p}\mathcal{N}_{q} = \ _{p}F_{q}$ and the derivatives of $_{p}F_{q}$ can be expressed in terms of the same $_{p}F_{q}$ with shifted parameter values \cite{as}. We then obtain for the factorial moments 
\begin{eqnarray}
& & _{p}n_{q}^{(k)}(x) \ = \ x^{k} \ 
\frac{(a_{1})_{k} \cdots (a_{p})_{k}}{(b_{1})_{k} \cdots (b_{q})_{k}} \times \nonumber \\ & & 
\frac{_{p}F_{q}(a_{1}+k,\ldots,a_{p}+k;b_{1}+k,\ldots,b_{q}+k; x)}
{_{p}F_{q}(a_{1},\ldots,a_{p};b_{1},\ldots,b_{q}; x)} \ . 
\end{eqnarray}
The mean photon number, 
$\bar{n} \equiv n^{(1)}$, is given by
\begin{equation}
_{p}\bar{n}_{q} = x \ 
{\textstyle \frac{a_{1} \cdots a_{p}}{b_{1} \cdots b_{q}} \ 
\frac{_{p}F_{q}(a_{1}+1,\ldots,a_{p}+1;b_{1}+1,\ldots,b_{q}+1;x)}
{_{p}F_{q}(a_{1},\ldots,a_{p};b_{1},\ldots,b_{q};x)}} \ ,
\end{equation}
and the Mandel parameter,  
$Q \equiv  - \bar{n} + n^{(2)} / \bar{n}$ \cite{mn, mw}, by
\begin{eqnarray}
& & _{p}Q_{q} = x \Big(-{\textstyle \frac{a_{1} \cdots a_{p}}{b_{1} \cdots b_{q}} \ 
\frac{_{p}F_{q}(a_{1}+1,\ldots,a_{p}+1;b_{1}+1,\ldots,b_{q}+1;x)}
{_{p}F_{q}(a_{1},\ldots,a_{p};b_{1},\ldots,b_{q};x)}} + \nonumber \\ & &  
{\textstyle \frac{(a_{1}+1) \cdots (a_{p}+1)}{(b_{1}+1) \cdots (b_{q}+1)} \ 
\frac{_{p}F_{q}(a_{1}+2,\ldots,a_{p}+2;b_{1}+2,\ldots,b_{q}+2;x)}
{_{p}F_{q}(a_{1}+1,\ldots,a_{p}+1;b_{1}+1,\ldots,b_{q}+1;x)}} \Big ).	
\end{eqnarray}
Positive, vanishing, or negative values of the Mandel parameter correspond to super-Poissonian, Poissonian, or sub-Poissonian statistics, respectively. The latter is a signature of nonclassical behaviour.
If in the above expressions we let $l$ $a$-parameters coalesce with $l$ $b$-parameters, we obtain the corresponding expressions for the lower-order states 
$\vert p-l;q-l;z \rangle$.

We now consider the photon number statistics of some particular GHCS. For each state we quote the photon number distribution, the mean photon number, and the Mandel parameter. We also show and discuss graphical representations for selected values 
of the parameters and compare them with those for the coherent states.

(a) We start with the coherent states $\vert \alpha \rangle$. 
Their photon number distribution is the Poisson distribution
$P_{\vert \alpha \rangle}(n) = (\vert \alpha \vert^{2n} / n!) \ 
{\text{e}}^{- \vert \alpha \vert^2}$, 
yielding $_{0}\bar{n}_{0} = \vert \alpha \vert^2$ and $_{0}Q_{0}=0$.

\begin{figure}
\includegraphics{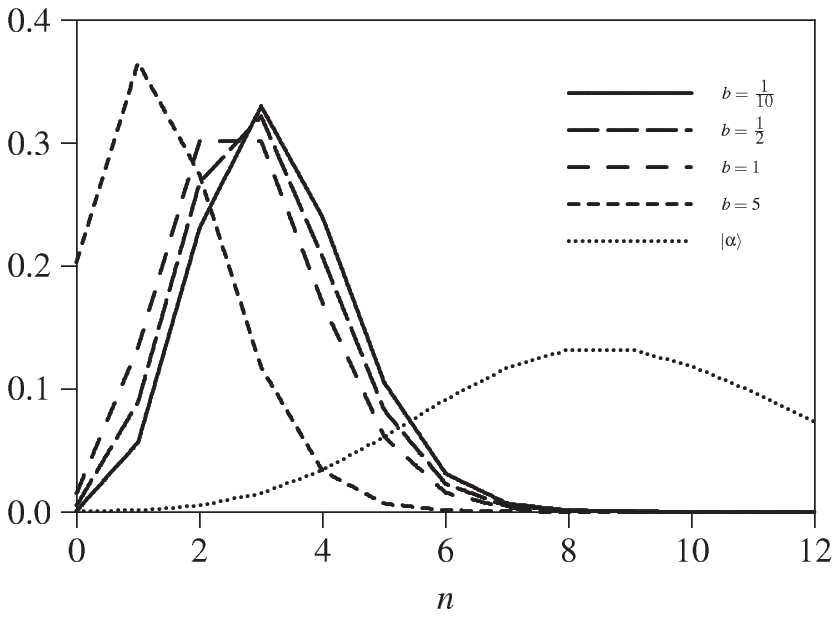}
\caption{\label{F1}Photon number distributions of the GHCS 
$\vert 0; 1; \alpha \rangle = \vert \cdot \thinspace ; b; \alpha \rangle _{(0;1)}$ 
and the CS $\vert \alpha \rangle$ for $\vert \alpha \vert =3$.}
\end{figure}

\begin{figure}
\includegraphics{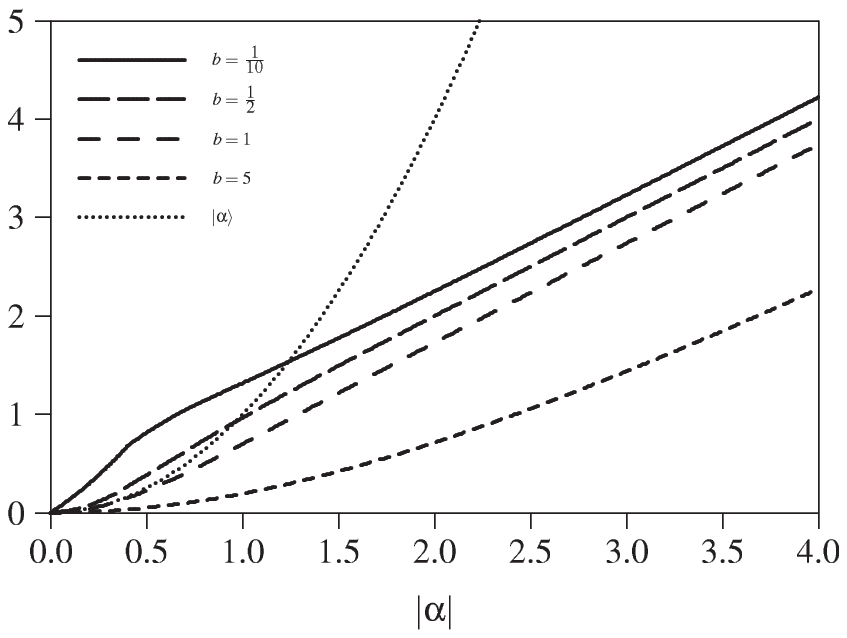}
\caption{\label{F2}Mean photon numbers of the GHCS 
$\vert 0; 1;\alpha \rangle = \vert \cdot \thinspace ; b; \alpha \rangle _{(0;1)}$ 
and the CS $\vert \alpha \rangle$ as a function of $\vert \alpha \vert$.}
\end{figure}

\begin{figure}
\includegraphics{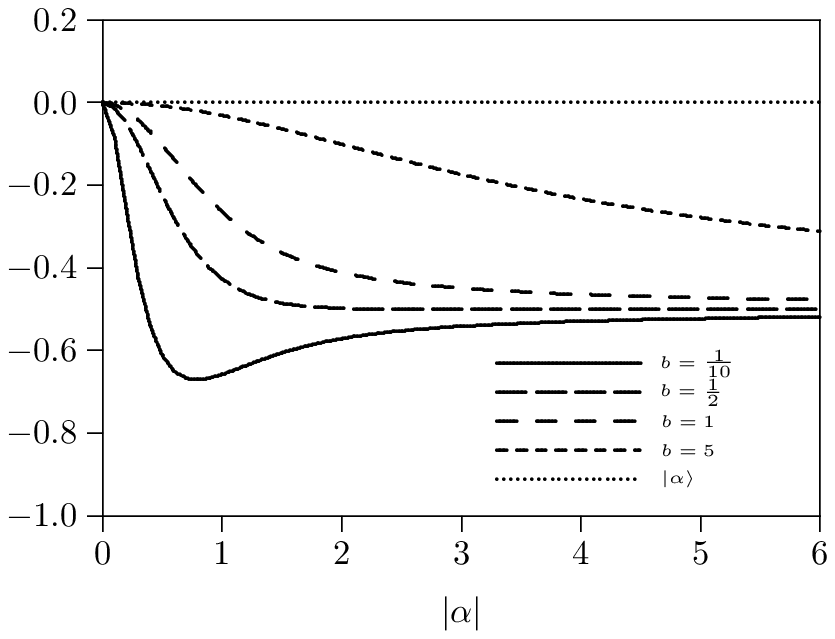}
\caption{\label{F3}Mandel parameters of the GHCS 
$\vert 0; 1; \alpha \rangle = \vert \cdot \thinspace ; b; \alpha \rangle _{(0;1)}$ 
and the CS $\vert \alpha \rangle$ as a function of $\vert\alpha\vert$.}
\end{figure}

(b) For the GHCS $\vert 0; 1; \alpha \rangle = 
\vert \cdot \thinspace ; b; \alpha \rangle _{(0;1)}$ we have:
\begin{equation}
P_{\vert 0; 1; \alpha \rangle}(n) \ = \ \frac{1}{n! \ \Gamma(b+n)} \ 
\frac{\vert \alpha \vert^{2n+b-1}}{I_{b-1}(2 \vert \alpha \vert)} ,
\label{G420}
\end{equation}
\begin{equation}
_{0}\bar{n}_{1} \ = \ \vert \alpha \vert \ \frac{I_{b}(2\vert
\alpha \vert)}{I_{b-1}(2\vert \alpha \vert)} , \label{G422}
\end{equation}
\begin{equation}
_{0}Q_{1} \ = \ \vert \alpha \vert \ \bigg( \frac{I_{b+1}(2 \vert
\alpha \vert)}{I_{b}(2 \vert \alpha \vert)} - \frac{I_{b}(2 \vert
\alpha \vert)}{I_{b-1}(2 \vert \alpha \vert)} \bigg) . \label{G424}
\end{equation}
The photon number distributions of the states $\vert 0; 1; \alpha \rangle$ in Fig.~\ref{F1} peak at low values of the photon number $n$, their mean photon numbers in Fig.~\ref{F2} increase for large enough $\vert\alpha\vert$ much slower than for the coherent state, and the negative values of their Mandel parameters in Fig.~\ref{F3} indicate nonclassical behaviour for all $\vert \alpha \vert$.

\begin{figure}
\includegraphics{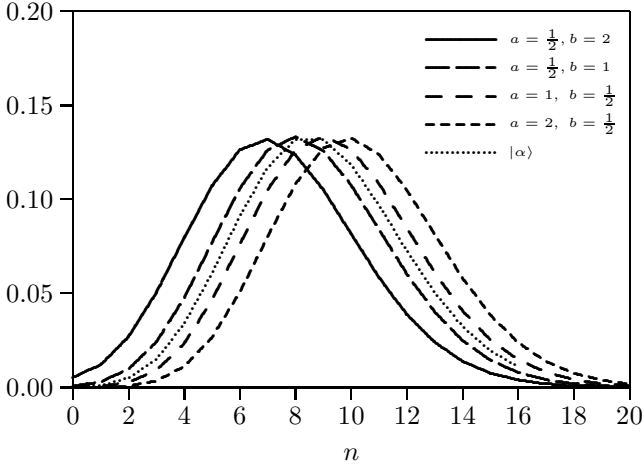}
\caption{\label{F4}Photon number distributions of the GHCS 
$\vert 1; 1; \alpha \rangle = \vert a; b; \alpha \rangle _{(1;1)}$ and the CS 
$\vert \alpha \rangle$ for $\vert \alpha \vert =3$.}
\end{figure}

\begin{figure}
\includegraphics{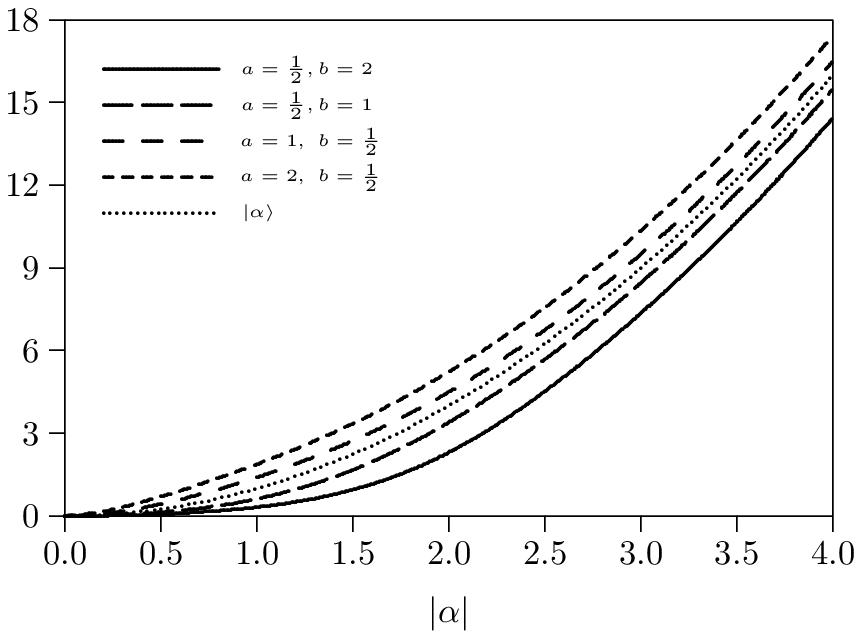}
\caption{\label{F5}Mean photon numbers of the GHCS 
$\vert 1; 1;\alpha \rangle = \vert a; b; \alpha \rangle _{(1;1)}$ and the CS $\vert \alpha \rangle$ as a function of $\vert \alpha \vert$.}
\end{figure}

\begin{figure}
\includegraphics{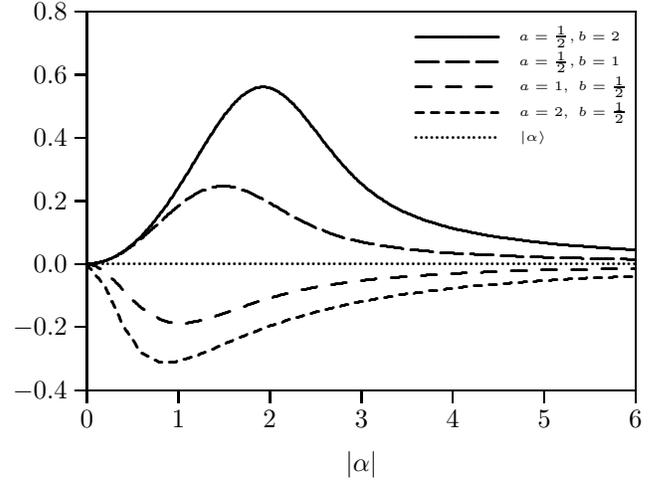}
\caption{\label{F6}Mandel parameters of the GHCS 
$\vert 1; 1; \alpha \rangle = \vert a; b; \alpha \rangle _{(1;1)}$ and the CS $\vert \alpha \rangle$ as a function of $\vert \alpha \vert$.}
\end{figure}

(c) For the GHCS $\vert 1; 1; \alpha \rangle = 
\vert a; b; \alpha \rangle _{(1;1)}$ we find
\begin{equation}
P_{\vert 1; 1; \alpha \rangle}(n) \ = \ \frac{(a)_{n}}{n! \
(b)_{n}} \ \frac{\vert \alpha \vert^{2n}}{ _{1}F_{1}(a; b; \vert
\alpha \vert^2)} , \label{G440}
\end{equation}
\begin{equation}
_{1}\bar{n}_{1} \ = \ \vert \alpha \vert^2 \ \frac{a}{b} \
\frac{_{1}F_{1}(a+1; b+1; \vert \alpha \vert^2)}{_{1}F_{1}(a; b;
\vert \alpha \vert^2)} , \label{G442}
\end{equation}
\begin{equation}
_{1}Q_{1} = -\ _{1}\bar{n}_{1} + 
{\textstyle \frac{a+1}{b+1} \ 
\frac{\vert \alpha \vert^2 \ _{1}F_{1}(a+2;b+2;\vert \alpha \vert^2)}
{_{1}F_{1}(a+1;b+1;\vert \alpha \vert^2)}} .
\label{G447}
\end{equation}
These relations reduce for $a=b$ to those of the coherent states. 
This remark is very helpful in understanding the behaviour of the photon number distributions in Fig.~\ref{F4}, the mean photon numbers in Fig.~\ref{F5}, and the Mandel parameters in Fig.~\ref{F6}. All curves coincide for $a=b$ with the CS-curves and depart from them as soon as $a$ becomes different from $b$. The deviations are on either side with respect to the CS-curves, depending on wether $a<b$ or $a>b$. As an example, the Mandel parameters in Fig.~\ref{F6} are positive (negative) for $a<b$ ($a>b$).

\begin{figure}
\includegraphics{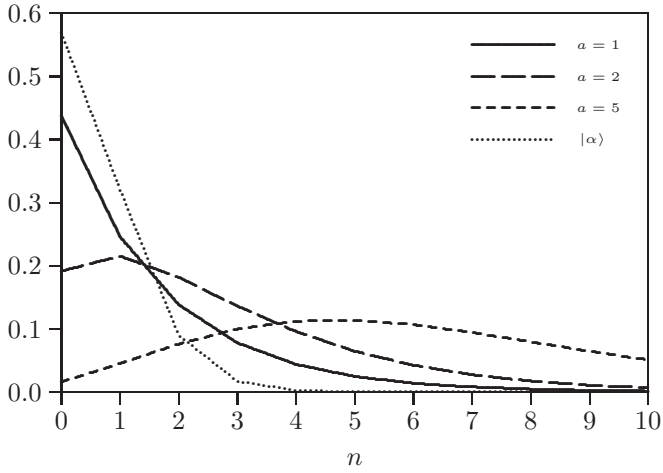}
\caption{\label{F7}Photon number distributions of the GHCS 
$\vert 1; 0; \epsilon \rangle = \vert a; \cdot \thinspace ; \epsilon \rangle _{(1;0)}$ and the CS $\vert \alpha \rangle$ for 
$\vert \epsilon \vert = \frac{3}{4} = \vert \alpha \vert$.}  
\end{figure}

(d) For the GHCS $\vert 1; 0; \epsilon \rangle = 
\vert a; \cdot \thinspace; \epsilon \rangle _{(1;0)}$ we have
\begin{equation}
P_{\vert 1; 0; \epsilon \rangle}(n) \ = \ \frac{(a)_{n}}{n!} \ (1 - \vert \epsilon \vert^2)^{a} \ \vert \epsilon \vert^{2n} \ , \label{G310}
\end{equation}
\begin{equation}
_{1}\bar{n}_{0} \ = \ a \  _{1}Q_{0} \ = \
\frac{a \ \vert \epsilon \vert^2}
{1 - \vert \epsilon \vert^2} \ , \label{G315}
\end{equation}
These expressions make sens for any $a>0$. Putting $a=2k$ and  $k=\frac{1}{2},1,\frac{3}{2},\ldots$ in Eq.~(\ref{G310}), we obtain the negative binomial distribution of the corresponding $SU(1,1)$ coherent states, also called negative binomial states for this reason \cite{vou1}.
The photon number distributions are shown in Fig.~\ref{F7} and compared with the Poisson distribution for the same value of the variable 
($\vert \epsilon \vert = \frac{3}{4}=\vert \alpha \vert$).
The mean photon number and the Mandel parameter have the same simple dependence on $\vert \epsilon \vert$. The Mandel parameter is independent of the parameter $a$ and always positive (classical behaviour). 

(e) For the GHCS $\vert 2; 1; \epsilon \rangle = 
\vert a_{1},a_{2}; b; \epsilon \rangle _{(2;1)}$ we obtain 
\begin{equation}
P_{\vert 2; 1; \epsilon \rangle}(n) = \frac{(a_{1})_{n} \
(a_{2})_{n}}{n! \ (b)_{n}} \frac{\vert \epsilon \vert^{2n}}{_{2}
F_{1}(a_{1},a_{2};b;\vert \epsilon \vert^2)} , \label{G360}
\end{equation}
\begin{equation}
_{2}\bar{n}_{1} = \vert \epsilon \vert^2 \
\frac{a_{1}a_{2}}{b} \ \frac{_{2}F_{1}(a_{1}+1,a_{2}+1;b+1; \vert
\epsilon \vert^2)}{_{2}F_{1}(a_{1}, a_{2}; b; \vert \epsilon
\vert^2)} , \label{G365}
\end{equation}
\begin{equation}
_{2}Q_{1} = - _{2}\bar{n}_{1} + \vert \epsilon \vert^2 
{\textstyle \frac{(a_{1}+1)(a_{2}+1)}{(b+1)} \ 
\frac{ _{2}F_{1}(a_{1}+2,a_{2}+2;b+2;\vert \epsilon \vert^2)}
{_{2}F_{1}(a_{1}+1,a_{2}+1;b+1;\vert \epsilon \vert^2)}} . 
\label{G370}
\end{equation}
For $b=a_{2}$ (or $b=a_{1}$) we recover the results of the previous
example (d). We refrain from graphical representations as there are too many parameters.

\section{Husimi and Husimi Phase Distributions} \label{S6}

In this section we consider the Husimi distributions ($Q$ functions) and the corresponding phase distributions for the generalized hypergeometric (coherent) states. We also comment on the Pegg-Barnett phase distributions.

The Husimi distribution for the GH(C)S $\vert p;q;z \rangle$ is given by the modulus squared of its wave function in the coherent state basis $\vert \alpha \rangle$
\begin{equation}
Q_{\vert p;q;z \rangle}(\alpha) \ = \ \frac{1}{\pi} \ \vert \langle \alpha \vert p;q;z \rangle \vert^2 ,  \label{G1110}
\end{equation}
where 
\begin{equation}
\langle \alpha \vert p; q; z \rangle \ = \ \Big[ \frac{{\text{e}}^{-\vert \alpha \vert^2}}{ _{p} {\mathcal N}_{q}(\vert z \vert^2)} \Big]^{\frac{1}{2}} \ \sum_{n=0}^{\infty} \frac{(\alpha^* z)^n}{\sqrt{n! \ _{p} \rho_{q}(n)}} \ . \label{G1131}
\end{equation}
It is positive, normalized to $1$ with measure 
$d^2 \alpha$ and provides a two-dimensional probability distribution over the complex $\alpha$-plane. Writing 
$\alpha = \vert \alpha \vert {\text{e}}^{i\theta}$ 
and integrating over the radial part $\vert \alpha \vert$, the Husimi phase distribution is obtained in the form \cite{ta,ch}
\begin{eqnarray}
\mathcal{P}_{\vert p;q;z \rangle}^{(Q)}(\theta) \ = \ \frac{1}{2
\pi} \sum_{n,n'=0}^{\infty} \ \langle n \vert p; q; z \rangle \ 
\langle n' \vert p; q; z \rangle^{*} \times \nonumber \\ 
\mathcal G^{(Q)}(n,n') \ {\text{e}}^{-i(n-n') \theta} , \  \label{G1145}
\end{eqnarray}
where the Fock components are given by Eq.~(\ref{G125}) and the newly introduced coefficients by
\begin{equation}
\mathcal G^{(Q)}(n,n') \ = \ 
\frac{\Gamma(\frac{n+n'}{2}+1)}{\sqrt{\Gamma(n+1) \ \Gamma(n'+1)}} \ . 
\label{G1420}
\end{equation}

In a different approach the phase states (\ref{G261}) can be used to define the phase distribution 
\begin{equation}
\mathcal{P}_{\vert p;q;z \rangle}^{(PB)}(\theta) \ = \ \vert \langle \theta \vert p;q;z \rangle \vert^2  \ . \label{G1422}
\end{equation}
It has the same form as Eq.~(\ref{G1145}), with 
$\mathcal G^{(Q)}(n,n')$ replaced by $\mathcal G^{(PB)}(n,n')=1$. 
Since the same result is obtained for the (differently motivated) Pegg-Barnett phase distribution \cite{pb}, we have used the upper index $(PB)$ in Eq.~(\ref{G1422}). Note that the dependence of the phase distributions on the phases is through the phase difference $(\theta - \varphi)$ only, where $\varphi$ is the phase of the complex variable ($z = \vert z \vert {\text{e}}^{i\varphi}$) defining the GH(C)S.

\begin{figure}
\includegraphics{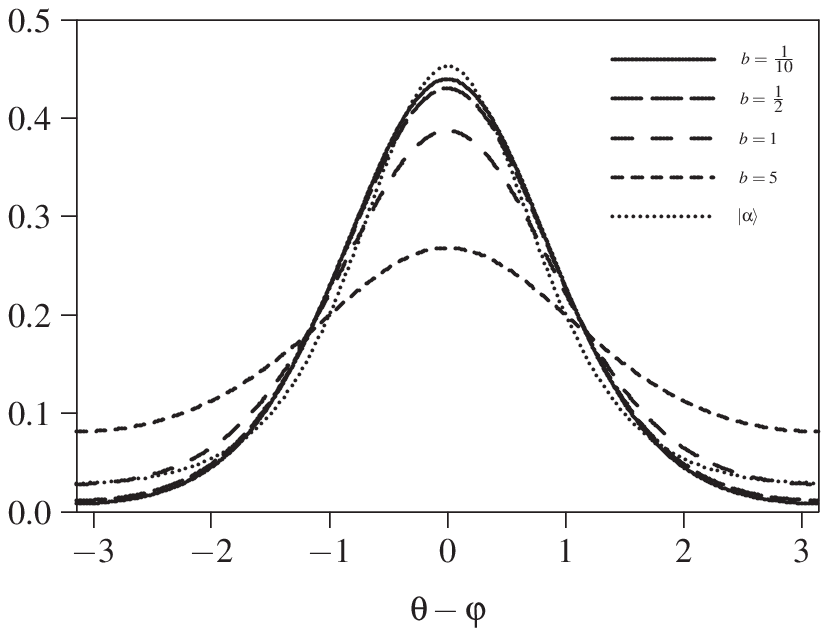}
\caption{\label{F8}Husimi phase distributions $\mathcal{P}_{\vert 0;1; \alpha  \rangle}^{(Q)}(\theta)$ of the GHCS 
$\vert 0;1; \alpha \rangle = \vert \cdot \thinspace ;b; \alpha \rangle _{(0;1)}$ and $\mathcal{P}_{\vert \alpha \rangle}^{(Q)}(\theta)$ of the CS $\vert \alpha \rangle$ for $\vert \alpha \vert = \frac{3}{4}$.}
\end{figure}

\begin{figure}
\includegraphics{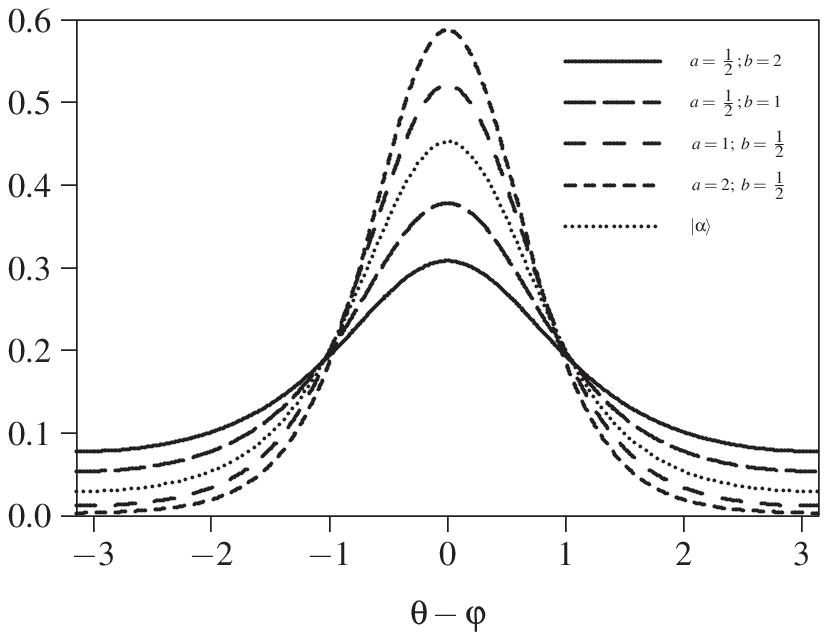}
\caption{\label{F9}Husimi phase distributions 
$\mathcal{P}_{\vert 1;1; \alpha \rangle}^{(Q)}(\theta)$
of the GHCS $\vert 1;1; \alpha \rangle = \vert a;b; \alpha \rangle _{(1;1)}$ 
and $\mathcal{P}_{\vert \alpha \rangle}^{(Q)}(\theta)$ of the CS $\vert \alpha \rangle$ for $\vert \alpha \vert = \frac{3}{4}$.}
\end{figure}

\begin{figure}
\includegraphics{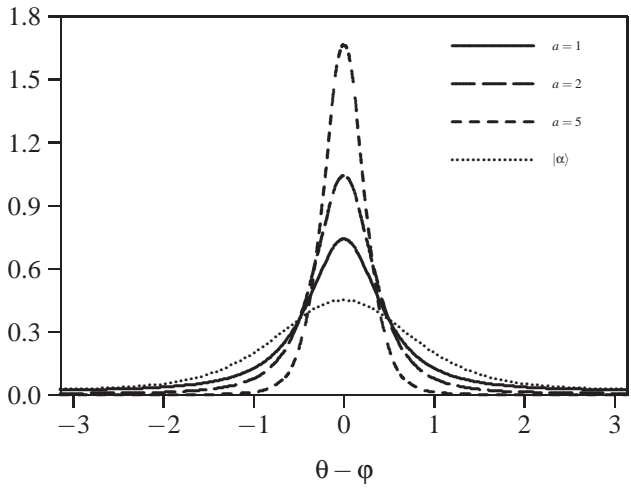}
\caption{\label{F10}Husimi phase distributions 
$\mathcal{P}_{\vert 1;0; \epsilon \rangle}^{(Q)}(\theta)$
of the GHCS $\vert 1;0; \epsilon \rangle = 
\vert a;\cdot \thinspace ; \epsilon \rangle _{(1;0)}$ and 
$\mathcal{P}_{\vert \alpha \rangle}^{(Q)}(\theta)$ of the CS $\vert \alpha \rangle$ for $\vert \epsilon \vert = \frac{3}{4} = \vert \alpha \vert $.}
\end{figure}

We display in Figures~\ref{F8}--\ref{F10} the Husimi phase distributions of the GHCS $\vert p; q; z \rangle$ for $(p;q) = (0;1), (1;1)$, and $(1;0)$, respectively, and for different values of the corresponding parameters. All distributions peak at $\theta = \varphi$. The heights of the peaks increase: with decreasing $b$ in Fig.~\ref{F8}, with decreasing $b$ and increasing $a$ in Fig.~\ref{F9}, and with increasing $a$ in Fig.~\ref{F10}. In Fig.~\ref{F9} the phase distributions 
$\mathcal{P}_{\vert 1; 1; \alpha \rangle}^{(Q)}(\theta)$ for $a=b$ and 
$\mathcal{P}_{\vert \alpha \rangle}^{(Q)}(\theta)$ would be identical; for $a>b$ ($a<b$) the heights of the peaks then move above (below) the CS peak. This is similar to the behaviour observed in the photon number distributions of the states 
$\vert 1; 1; \alpha \rangle$.

The Pegg-Barnett phase distributions look qualitatively similar, so we do not display them explicitly. The peaks at $\theta = \varphi$ are always higher 
(due to $\mathcal{G}^{(PB)}(n,n') \ge \mathcal{G}^{(Q)}(n,n')$) 
and consequently the distributions are narrower in order to preserve the normalization.

\section{Generalized Hypergeometric Husimi and Generalized Hypergeometric Husimi Phase Distributions} \label{S7}

By analogy with the conventional Husimi distribution we define in this section the generalized hypergeometric Husimi distribution of a normalized state $\vert \psi \rangle$  by the modulus squared of its wave function in the GHCS basis, Eq.~(\ref{G299}):
\begin{eqnarray}
Q_{\vert \psi \rangle}^{(p;q)}(z) &\equiv& 
Q_{\vert \psi \rangle}^{(p;q)}(a_{1},\ldots, a_{p};b_{1},\ldots, b_{q};z) \nonumber \\ &=& \frac{1}{\pi} \ _{p}w_{q}(\vert z \vert^2) \ \langle p; q; z \vert \psi \rangle \langle \psi \vert p; q; z \rangle . \ \  
\label{G1120}
\end{eqnarray}
It provides a two-dimensional true probability distribution over the complex $z$-plane for $p<q+1$ and over the unit $z$-disk for $p=q+1$. The corresponding normalization condition
\begin{equation}
\int\limits d^2 z \ Q_{\vert \psi \rangle}^{(p;q)}(z) \ = \ 1 
\label{G1130}
\end{equation}
is a consequence of the resolution of unity, Eq.~(\ref{G280}). It is therefore mandatory for the state $\vert p; q; z \rangle$ in the definition (\ref{G1120}) to be a GHCS; the state $\vert \psi \rangle$ can be any normalized state, in particular a GHS or GHCS. For later reference, we call the states $\vert \psi \rangle$ and $\vert p; q; z \rangle$ the signal and analyzing states, respectively. In defining generalized hypergeometric Husimi distributions we have formally replaced the analyzing CS in the usual Husimi distribution by an analyzing GHCS and have included the weight function into the distribution rather than into the integration measure. The conventional Husimi distribution is then recovered for $p=q=0$, i.e. 
$Q_{\vert \psi \rangle}(\alpha) \ = \ Q_{\vert \psi \rangle}^{(0;0)}(\alpha)$.

The generalized hypergeometric Husimi distributions can be used to define corresponding phase distributions by integrating over the modulus of the complex variable $z = \vert z \vert {\text{e}}^{i\theta}$ ($x = \vert z \vert^2$):
\begin{equation}
\mathcal{P}_{\vert \psi \rangle}^{(p;q)}(\theta) \ \equiv \ \frac{1}{2} \ 
\int\limits_{0}^{R} dx \ Q_{\vert \psi \rangle}^{(p;q)}
(\sqrt{x} \ {\text{e}}^{i\theta}) \ ,  
\label{G1140}
\end{equation}
where $R=\infty$ ($R=1$) for the GHCS on the plane (disk).
Performing the integral with the help of Eq.~(\ref{G281}), obviously extendable to arbitrary (here halfinteger) $n$, the generalized hypergeometric Husimi phase distributions are obtained in the form 
\begin{equation}
\mathcal{P}_{\vert \psi \rangle}^{(p;q)}(\theta) \ = \ 
\frac{1}{2 \pi} \sum_{n,n'=0}^{\infty} \ \psi_{n} \ \psi_{n'}^{*} \ 
\mathcal G^{(p;q)}(n,n') \ {\text{e}}^{-i(n-n') \theta} \ .
\label{G1150}
\end{equation}
They are normalized according to 
\begin{equation}
\int\limits d \theta \ \mathcal{P}_{\vert \psi \rangle}^{(p;q)}(\theta) \ = \ 1 \ , 
\label{G1155}
\end{equation}
where the integration is over the $2 \pi$-interval chosen for the phase $\theta$, usually $[- \pi, \pi]$ or $[0, 2 \pi]$.

The phase distributions (\ref{G1150}) and (\ref{G1145}) have the same formal structure, only the coefficients are different. Those appearing in Eq.~(\ref{G1150}) are given by
\begin{equation}
\mathcal G^{(p;q)}(n,n') \ = \
\frac{_{p}\rho_{q}(\frac{n+n'}{2})}{\sqrt{_{p}\rho_{q}(n) \
_{p}\rho_{q}(n')}} 
\label{G1160}
\end{equation}
in terms of the moments $_{p}\rho_{q}(n)$ from Eq.~(\ref{G230}). They are symmetric in $n$ and $n'$ and normalized to $1$ for $n=n'$, thus insuring the normalization condition (\ref{G1155}). These coefficients encode the information on the analyzing state $\vert p; q; z \rangle$. The information on the signal state $\vert \psi \rangle$ is contained solely in the product $\psi_{n} \ \psi_{n'}^{*}$, where $\psi_{n}=\langle n \vert \psi \rangle$. If the system is described by a density operator $\hat \rho$, this product should be replaced by the density matrix elements in the Fock basis, 
$\rho _{n,n'} = \langle n \vert \hat \rho \vert n' \rangle$; correspondingly, 
$\hat \rho$ should appear instead of $\vert \psi \rangle \langle \psi \vert$ in Eq.~(\ref{G1120}). If the dependence on the parameters is needed, we shall write 
$\mathcal{P}_{\vert \psi \rangle}^{(p;q)}(a_{1},\ldots, a_{p};b_{1},\ldots, b_{q}; \theta)$ for the phase distributions and 
$\mathcal G^{(p;q)}(a_{1},\ldots, a_{p};b_{1},\ldots, b_{q};n,n')$ for the coefficients. The latter can be written also in factorized form 
\begin{equation}
\mathcal G^{(p;q)}(n,n') = \mathcal G^{(Q)}(n,n') \   
\frac{\prod_{j=1}^{q} \mathcal G^{(b_{j})}(n,n')}
{\prod_{i=1}^{p} \mathcal G^{(a_{i})}(n,n')} \ ,
\label{G1170}
\end{equation}
with the one-parameter coefficients given by
\begin{equation}
\mathcal G^{(a)}(n,n') \ = \ 
\frac{\Gamma(\frac{n+n'}{2}+a)}{\sqrt{\Gamma(n+a) \ \Gamma(n'+a)}} \ . 
\label{G1180}
\end{equation}
Comparing the product structures of $_{p}\rho_{q}(n)$ and 
$\mathcal G^{(p;q)}(n,n')$, Eqs.~(\ref{G230}) and (\ref{G1170}), it follows that each Pochhammer symbol $(a)_{n}$ contributes a coefficient $\mathcal G^{(a)}(n,n')$; 
the factor $\mathcal G^{(Q)}(n,n') = \mathcal G^{(1)}(n,n')$ comes from 
$\Gamma (n+1) = (1)_{n}$ according to the same rule. 
If in Eq.~(\ref{G1170}) $p$ or $q$ or both are zero, the corresponding product(s) should be replaced by $1$. Thus 
$\mathcal G^{(0;0)}(n,n') = \mathcal G^{(Q)}(n,n')$, 
so that the conventional Husimi phase distribution is given by
$\mathcal{P}_{\vert \psi \rangle}^{(Q)}(\theta) = 
\mathcal{P}_{\vert \psi \rangle}^{(0;0)}(\theta)$.
Similarly, the Pegg-Barnett phase distribution is given by
$\mathcal{P}_{\vert \psi \rangle}^{(PB)}(\theta) = 
\mathcal{P}_{\vert \psi \rangle}^{(1;0)}(a;\cdot \thinspace ; \theta) \mid _{a=1}$, since 
$\mathcal G^{(1;0)}(a;\cdot \thinspace ;n,n') = 
\mathcal G^{(Q)}(n,n') / \mathcal G^{(a)}(n,n')$ 
equals $1$ for $a = 1$. More generally, the generalized hypergeometric Husimi and corresponding phase distributions of order $(p-l;q-l)$ result from those of order $(p;q)$ by letting $l$ numerator parameters coalesce with $l$ denominator parameters.

We now consider some generalized hypergeometric Husimi phase distributions for the Fock and coherent states. For any Fock state $\vert N \rangle$ we have $\psi_{n} = \delta_{n,N}$ and the phase distributions (\ref{G1150}) become uniform with value $1 /(2\pi)$, as expected.

\begin{figure}
\includegraphics{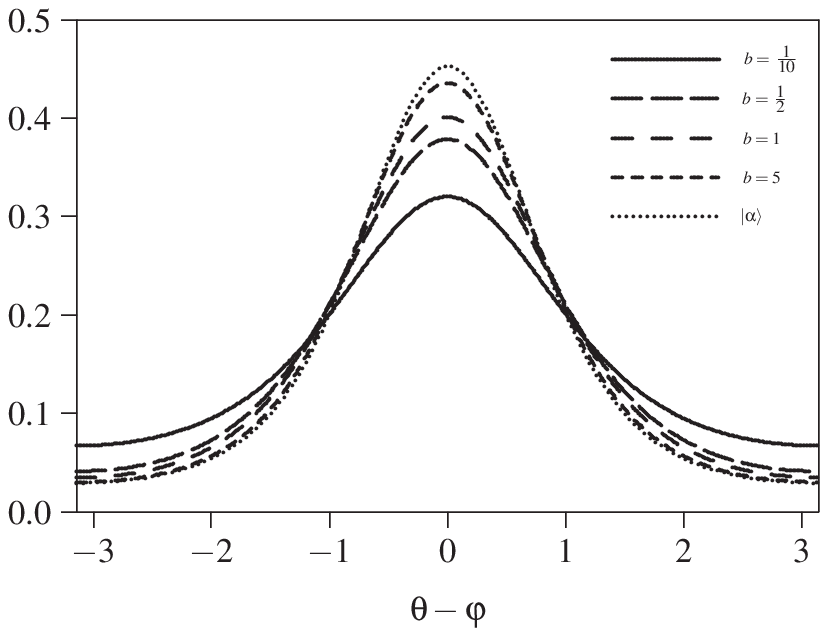}
\caption{\label{F11}Generalized hypergeometric Husimi phase distributions $\mathcal{P}_{\vert \alpha \rangle}^{(0;1)}(\cdot \thinspace ; b; \theta)$ and Husimi phase distribution $\mathcal{P}_{\vert \alpha \rangle}^{(Q)}(\theta)$ of the CS $\vert \alpha \rangle$ for $\vert \alpha \vert = \frac{3}{4}$.}
\end{figure}

\begin{figure}
\includegraphics{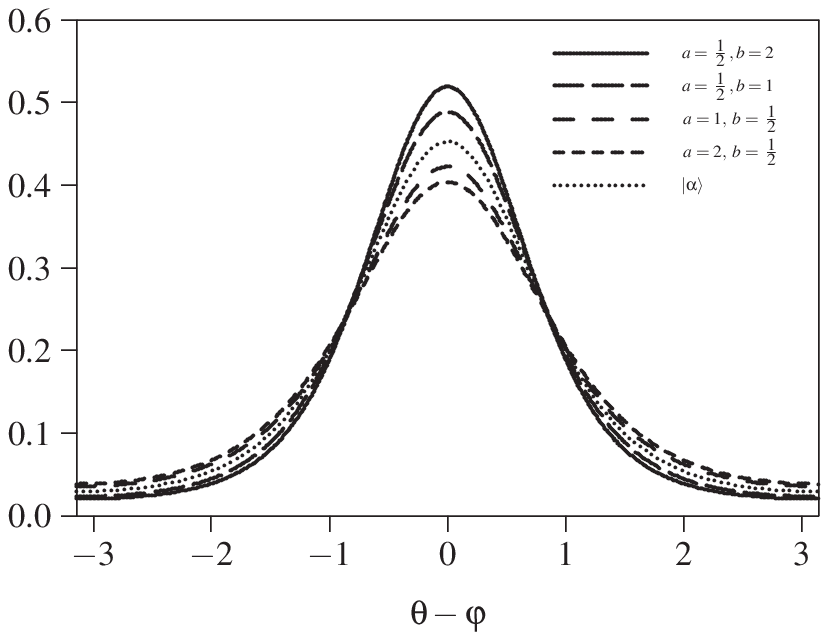}
\caption{\label{F12}Generalized hypergeometric Husimi phase distributions 
$\mathcal{P}_{\vert \alpha \rangle}^{(1;1)}(a; b; \theta)$ and Husimi phase distribution
$\mathcal{P}_{\vert \alpha \rangle}^{(Q)}(\theta)$ of the CS 
$\vert \alpha \rangle$ for $\vert \alpha \vert = \frac{3}{4}$.}
\end{figure}

\begin{figure}
\includegraphics{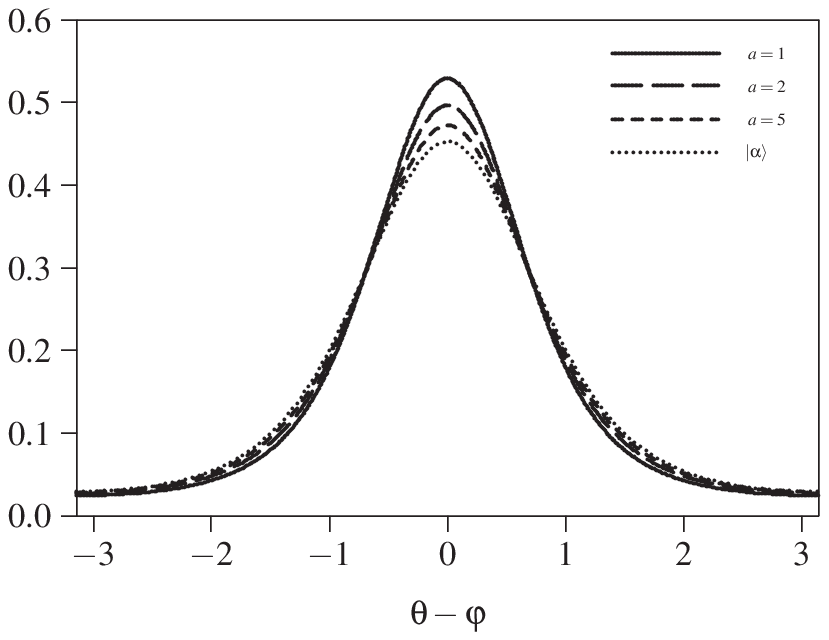}
\caption{\label{F13}Generalized hypergeometric Husimi phase distributions 
$\mathcal{P}_{\vert \alpha \rangle}^{(1;0)}(a; \cdot \thinspace ; \theta)$ and Husimi phase distribution 
$\mathcal{P}_{\vert \alpha \rangle}^{(Q)}(\theta)$ of the CS 
$\vert \alpha \rangle$ for $\vert \alpha \vert = \frac{3}{4}$.}
\end{figure}

For the coherent states $\vert \alpha \rangle$ with 
$\alpha = \vert \alpha \vert {\text{e}}^{i \varphi}$ and 
$\psi_{n} = \text{e}^{- \vert \alpha \vert^2 / 2} \ \vert \alpha \vert^n \ {\text{e}}^{i n \varphi} / \sqrt{n!}$, 
the resulting phase distributions (\ref{G1150}) depend only on the phase difference ($\theta - \varphi$). We display in Figures~\ref{F11}--\ref{F13} the generalized hypergeometric Husimi phase distributions of a coherent state for $(p;q) = (0;1), (1;1)$, and $(1;0)$, respectively. All distributions peak at the coherent phase $\theta = \varphi$. The heights of the peaks decrease: with decreasing $b$ in Fig.~\ref{F11}, with decreasing $b$ and increasing $a$ in Fig.~\ref{F12}, and with increasing $a$ in Fig.~\ref{F13}. In Fig.~\ref{F12} the phase distribution $\mathcal{P}_{\vert \alpha \rangle}^{(1;1)}(a;b;\theta)$ for $a=b$ and the Husimi phase distribution $\mathcal{P}_{\vert \alpha \rangle}^{(Q)}(\theta)$ would yield the same results; for $a<b$ ($a>b$) the heights of the peaks then move above (below) the peak of 
$\mathcal{P}_{\vert \alpha \rangle}^{(Q)}(\theta)$. 
This behaviour is similar to the one observed in the conventional Husimi phase distributions of the states $\vert 1; 1; \alpha \rangle$ as well as in their photon number distributions.

The phase distributions shown in Figures~\ref{F11}--\ref{F13} are ``dual'' to those in Figures~\ref{F8}--\ref{F10} in the sense that the signal and analysing states have been interchanged. The signal (analyzing) state is a GHCS (CS) in Figures~\ref{F8}--\ref{F10}, and a CS (GHCS) in Figures~\ref{F11}--\ref{F13}. In the first case we deal with the usual Husimi phase distribution 
$\mathcal{P}_{\vert p; q; z \rangle}^{(0;0)}$ 
of the GHCS $\vert p; q; z \rangle$, in the second case with the generalized hypergeometric Husimi phase distribution 
$\mathcal{P}_{\vert \alpha \rangle}^{(p;q)}$ of the conventional CS $\vert \alpha \rangle$. The two sets of figures show an opposite behaviour in the sense that for the same variation of the parameters the heights of the peaks increase in one set and decrease in the other set.

The phase distributions showing this dual behaviour originate from generalized hypergeometric Husimi distributions of the type 
$Q_{\vert \tilde p; \tilde q; \tilde z \rangle}^{(p;q)}(z)$, 
where both the signal ($\vert \tilde p; \tilde q; \tilde z \rangle$) and the analysing ($\vert p;q;z \rangle$) states are GHCS and therefore interchangeable. 
We call the generalized hypergeometric Husimi distribution 
$Q_{\vert \tilde p; \tilde q; \tilde z \rangle}^{(p;q)}(a_{1},\ldots, a_{p}; b_{1},\ldots, b_{q};z)$ and the one with interchanged states,
$Q_{\vert p; q; z \rangle}^{(\tilde p;\tilde q)}(\tilde a_{1},\ldots, \tilde a_{p}; \tilde b_{1},\ldots, \tilde b_{q};\tilde z)$, 
dual to each other and extend this definition to the corresponding phase distributions, 
$\mathcal{P}_{\vert \tilde p; \tilde q; \tilde z \rangle}^{(p;q)}(\theta)$ and
$\mathcal{P}_{\vert p; q; z \rangle}^{(\tilde p; \tilde q)}(\theta)$. 
Dual distributions are nontrivial due to the nonorthogonality of the GHCS and this is true even if for $p=\tilde p$ and $q=\tilde q$ the two sets of tilded and untilded parameters are identical. In this case the self-dual distributions $Q_{\vert p; q; \tilde z \rangle}^{(p;q)}(z)$ are determined by the scalar product in Eq.~(\ref{G237}), i.e.
\begin{equation}
Q_{\vert p; q; \tilde z \rangle}^{(p;q)}(z) \ = \ 
\frac{1}{\pi} \ \frac{_{p}w_{q}(\vert z \vert^2) \ 
\vert _{p}\mathcal{N}_{q}(z^{*} \tilde z) \vert^2}
{_{p} \mathcal{N}_{q}(\vert z \vert^2) \  
_{p} \mathcal{N}_{q}(\vert \tilde z \vert^2)} \ .
\end{equation}
Self-dual distributions actually provide a measure of the nonorthogonality of the GHCS with identical parameter sets. The simplest self-dual distribution is the usual Husimi distribution for a conventional coherent state, given by the well known overlap of the two coherent states: 
$Q_{\vert \tilde \alpha \rangle}(\alpha) = 
\frac{1}{\pi}\vert \langle \alpha \vert \tilde \alpha \rangle \vert^2 = 
Q_{\vert \alpha \rangle}(\tilde \alpha)$. 
The corresponding self-dual phase distribution is 
$\mathcal{P}_{\vert \alpha \rangle}^{(Q)}(\theta)$, 
also shown in Figures~\ref{F8}--\ref{F13} for comparison.

\section{Summary} \label{S8}

In this paper we have introduced a large class of holomorphic quantum states
$\vert a_{1},\ldots, a_{p}; b_{1},\ldots, b_{q}; z \rangle$, 
depending on a complex variable $z$ and two sets of parameters, 
$(a_{1},\ldots, a_{p})$ and $(b_{1},\ldots, b_{q})$. 
We called them generalized hypergeometric states (GHS) as their normalization function is given by the generalized hypergeometric function 
$\ _{p} F_{q}(a_{1},\ldots, a_{p}; b_{1},\ldots, b_{q}; \vert z \vert^2)$. 
Depending on the domain of convergence of the latter, the GHS are defined on the whole $z$-plane ($\vert z \vert < \infty$), the open unit disk ($\vert z \vert < 1$), or the unit circle ($\vert z \vert = 1$). Accordingly, these states may be considered as generalizations of the usual coherent states, the coherent phase states, and the phase states. Those GHS yielding a resolution of unity with a positive weight function define the subclass of generalized hypergeometric 
coherent states (GHCS). We have determined the corresponding weight functions by Mellin transform techniques, but their positivity must be checked separately. The GHS on the plane (unit disk and circle) are eigenstates with eigenvalue $z$ of suitably defined lowering operators 
$_{p} \hat A_{q}$ ($_{p} \hat E_{q}$); the latter may be considered as generalizations of $\hat a$ ($\hat E$), the usual annihilation operator of the harmonic oscillator (the Susskind-Glogower exponential phase operator).

To reveal the physical content of the GH(C)S we have studied their photon number statistics (photon number distribution, mean photon number, Mandel parameter) and phase properties (Husimi and Pegg-Barnett phase distributions). Numerical results were displayed for particular states and parameter values.

The GHCS on the plane (unit disk) can be used to define new analytic representations of arbitrary quantum states, which are elements of corresponding Bargmann (Hardy) spaces. They can be used also to define generalized hypergeometric Husimi functions over the complex plane (unit disk) and corresponding phase distributions. Of particular interest are the generalized hypergeometric Husimi functions when considered for another GHCS, as then the two GHCS can be interchanged (dual distributions). The corresponding phase distributions show a correlated behaviour. Self-dual distributions are nontrivial and provide a measure of the nonorthogonality of the GHCS.

\end{document}